
%
%
\input phyzzx
\nonstopmode
\twelvepoint
\nopubblock
\overfullrule=0pt
\tolerance=5000

\line{\hfill }
\line{\hfill PUPT-1347, IASSNS-HEP-92/60}
\line{\hfill October 1992}

\titlepage
\title{ Static and Dynamic Critical Phenomena at a Second Order
QCD Phase Transition}

\author{Krishna Rajagopal\foot{Research supported in part by a Natural
Sciences and Engineering Research Council of Canada 1967 Fellowship.~~~
RAJAGOPAL@PUPGG.PRINCETON.EDU}}
\vskip .2cm
\centerline{{\it Department of Physics }}
\centerline{{\it Joseph Henry Laboratories }}
\centerline{{\it Princeton University }}
\centerline{{\it Princeton, N.J. 08544 }}

\author{Frank Wilczek\foot{Research supported in part by DOE grant
DE-FG02-90ER40542.~~~WILCZEK@IASSNS.BITNET}}
\vskip.2cm
\centerline{{\it School of Natural Sciences}}
\centerline{{\it Institute for Advanced Study}}
\centerline{{\it Olden Lane}}
\centerline{{\it Princeton, N.J. 08540}}
\endpage

\abstract{ In QCD with two flavors of massless quarks, the
chiral phase transition is
plausibly in the same universality class as the
classical four component Heisenberg antiferromagnet.
Therefore, renormalization group techniques developed in the
study of phase transitions can be applied to calculate
the critical exponents
which characterize the scaling behaviour of
universal quantities near the critical point.  This approach
to the QCD phase transition has implications both for lattice gauge
theory and for heavy ion collisions.  Future lattice simulations with
longer correlation lengths will be able to measure the various exponents
and the equation of state for the order parameter as a function of
temperature and quark mass which we describe.  In a heavy ion
collision, the consequence of a long correlation length would be large
fluctuations in the number ratio of neutral to charged pions.  Unfortunately,
we show that this phenomenon will not occur if the plasma stays close to
equilibrium as it cools.  If the transition is far out of equilibrium
and can be modelled as a quench, it is possible that large volumes
of the plasma with the pion field correlated will develop, with dramatic
phenomenological consequences. }

\endpage

\REF\wilczek{F. Wilczek, {\it Int. J. Mod. Phys.} {\bf A7}, 3911 (1992).
This elaborates earlier work of R. Pisarski and F. Wilczek, {\it Phys. Rev.}
{\bf D29}, 338 (1984).}

\REF\thooft{ G. 't Hooft, {\it Phys. Reports}  {\bf 142}, 357 (1986).}

\REF\amitma{See, for example, {\it Modern Theory of Critical Phenomena},
S.-K. Ma (Benjamin/Cummings, 1976) and {\it Field Theory, the Renormalization
Group, and Critical Phenomena}, D. J. Amit (World Scientific, 1984).}

\REF\gm{M. Gell-Mann and M. Levy, {\it Nuovo Cimento} {\bf 16}, 705 (1960).}

\REF\ginsparg{P. Ginsparg, {\it Nucl. Phys.} {\bf B170}, 388 (1980),
 and references therein.}

\REF\gocksch{A. Gocksch, {\it Phys. Rev. Lett.} {\bf 67}, 1701 (1991).}

\REF\bmn{G. Baker, B. Nickel and D. Meiron, {\it Phys. Rev.} {\bf B17},
1365 (1978), and ``Compilation of 2-pt. and 4-pt. graphs for continuous
spin models'', unpublished, University of Guelph report (1977).}

\REF\kocic{A. Koci\'c, J. B. Kogut, and M.-P. Lombardo, Illinois preprint
ILL-TH-92-18, 1992.}

\REF\bww{E. Br\'ezin, D. J. Wallace and K. G. Wilson, {\it Phys. Rev.} {\bf B7}
232 (1973).}

\REF\bw{E. Br\'ezin and D. J. Wallace, {\it Phys. Rev.} {\bf B7} 1967 (1973).}

\REF\wz{D. J. Wallace and R. K. P. Zia, {\it Phys. Rev.} {\bf B12} 5340
(1975).}

\REF\hk{T. Hatsuda and T. Kunihiro, {\it Phys. Rev. Lett.} {\bf 55} 158 (1985)
and references therein.}

\REF\pokorski{{\it Gauge Field Theories}, S. Pokorski (Cambridge, 1987).}

\REF\columbia{F. R. Brown {\it et al.}, {\it Phys. Rev. Lett.} {\bf 65} 2491
(1990).}

\REF\lawrie{I. Lawrie and S. Sarbach in {\it Phase Transitions and Critical
Phenomena} {\bf 9}, 1 (1984), ed. C. Domb and J. Lebowitz (Academic
Press).}

\REF\hh{P. C. Hohenberg and B. I. Halperin, {\it Rev. Mod. Phys.} {\bf 49},
435 (1977).  B.~I.~Halperin, P. C. Hohenberg, and E. D. Siggia,
{\it Phys. Rev.} {\bf B13}, 1299 (1976).}

\REF\hhm{B. I. Halperin, P. C. Hohenberg, and S.-K. Ma, {\it Phys. Rev.}
{\bf B10}, 139 (1974).}

\REF\bernard{C. Bernard {\it et al.}, {\it Phys. Rev.} {\bf D45}, 3854 (1992).}

\REF\boyd{G. Boyd {\it et al.}, {\it Nucl. Phys.} {\bf B376}, 199 (1992).}

\REF\wilsonquarks{C. Bernard {\it et al.}, {\it QCD Thermodynamics With
Two Flavors of Wilson Quarks at $N_t =6$}, Santa Barbara preprint
UCSBTH-92-30.}

\REF\bak{P. Bak and  D. Mukamel, {\it Phys. Rev.} {\bf B13}, 5086 (1976).}

\REF\barak{Z. Barak and M. B. Walker, {\it Phys. Rev.} {\bf B25}, 1969 (1982).}

\REF\bjorken{J. D. Bjorken, {\it Phys. Rev.} {\bf D27}, 140 (1983).}

\REF\anselm{A. A. Anselm and M. G. Ryskin, {\it Phys. Lett.} {\bf B266}, 482
(1991).}

\REF\blaizot{J.-P. Blaizot and A. Krzywicki, {\it Phys. Rev.} {\bf D46}, 246
(1992).}

\REF\leutwyler{J. Gasser and H. Leutwyler, {\it Phys. Lett.} {\bf B184},
83 (1987).}

\REF\loewe{C. Contreras and M. Loewe, {\it Int. J. Mod. Phys.} {\bf A5},
2297 (1990).}

\REF\kaonmass{H. Leutwyler, {\it Nucl. Phys.} {\bf B337}, 108 (1990).}

\REF\centauro{C. M. G. Lattes, Y. Fujimoto, and S. Hasegawa, {\it Phys. Rept.}
{\bf 65}, 151 (1980).}

\REF\bray{A. J. Bray, {\it Phys. Rev.} {\bf B41}, 6724 (1990);  T. J. Newman,
A. J. Bray, and M. A. Moore, {\it Phys. Rev.} {\bf B42}, 4514 (1990); and
references therein.}

\REF\turgel{N. Turok and D. N. Spergel, {\it Phys. Rev. Lett.} {\bf 66}, 3093
(1991).}

\REF\inprogress{K. Rajagopal and F. Wilczek, work in progress.}

\chapter{Introduction}

The QCD phase transition is of interest from several different points of
view.  First, there can be no doubt that it occurred in the early universe.
Second, it is reasonable to hope that in a heavy ion collision of
sufficiently high energy, a small region of the high temperature phase
is created which then cools through the phase transition.
Third, lattice gauge theory is well suited to calculating the equilibrium
properties of QCD at high temperatures.
{}From all these perspectives, it is important to learn as much as
can be learned analytically about the phase transition, relying as much
as possible only on fundamental symmetries and universality arguments and
as little as possible on specific assumptions and models.

In a previous paper [\wilczek ], one of us (F. W.)
emphasized that in the chiral limit
where there are two species of quarks with zero current algebra mass, the
order parameter for the chiral phase transition has the same symmetry as
the magnetization of a four component Heisenberg magnet, which
has a second order phase transition.  In this case,
and indeed for any number of quark species except zero, there is no
order parameter for a confinement/deconfinement phase
transition.  Thus in the $m_u=m_d=0$ limit
the universal characteristics of the QCD phase transition ({\it i.e.} those
characteristics determined by the modes which develop long correlation
lengths at the phase transition) are the same as those of the $N=4$
Heisenberg magnet.

In this paper we further explore the consequences of this approach
to the QCD phase transition.  In the following section, we review the
scenario described in [\wilczek ], and establish a dictionary between QCD and
the magnetic system.  In order to make the present paper self-contained,
we also include
in this section
many of the results from [\wilczek ].  In section 3, we discuss
the behaviour of the pion and sigma masses at the transition.  In section
4, we discuss the $\rho$ and $A_1$ mesons.  In section 5, we consider how the
 strange
quark affects the phase transition.
In section 6, we go beyond the static
critical phenomena of the earlier sections and discuss the dynamics
of the appropriate universality class.  In section 7, we discuss the
implications of all this for cosmology and lattice gauge theory,
and for heavy ion collisions under the assumption that the plasma
remains close to thermal equilibrium through the phase transition.
In section 8, we consider what phenomena we can expect in heavy
ion collisions if the system gets far out of thermal equilibrium and
can be modelled as a quench.   Finally, in the last section we
summarize and conclude.

\chapter{QCD and the $O(4)$ Magnet}

As discussed in [\wilczek ], the physics of the QCD phase transition is
qualitatively different in the cases of zero, one, two, or three or more
flavors of quarks.
In this section we consider QCD with two species
of quarks.  (An analysis similar to the one which follows leads to the
conclusion that for three or more flavors of massless quarks, the
chiral phase transition is first order.  See [\wilczek ] for details.)
If there are two flavors of massless quarks, the lagrangian is symmetric under
global chiral transformations in the group $SU(2)_L \times SU(2)_R \times
U(1)_{L+R}$ of independent special unitary transformations of the left and
right handed quark fields, and a vector $U(1)$ transformation which
corresponds to baryon number symmetry.  (The axial $U(1)$ which would make
the symmetry group into $U(2) \times U(2)$ is a symmetry of the classical
theory, but not of the quantum theory [\thooft ].)
This chiral symmetry
breaks spontaneously down to $SU(2)_{L+R} \times U(1)_{L+R}$
at low temperatures; and is restored at sufficiently high temperatures.
The order parameter for this phase transition is the expectation value
of the quark bilinear
$$
{\cal M}^i_j ~\equiv ~ \langle  {\bar q_L}^i {q_R}_j \rangle
\eqn\ab
$$
which breaks the symmetry when it acquires a non-zero value below some
critical temperature $T_c$.

In order to describe a second-order transition
quantitatively, we must find a tractable model in the
same universality class.  For the chiral order parameter
\ab\ the relevant symmetries are independent unitary transformations
of the left- and right-handed quark fields, under which
$$
{\cal M}~\rightarrow~ U^\dagger {\cal M} V ~.
\eqn\ba
$$
These transformations generate an
$U(2)_L\times U(2)_R $ symmetry, which is not quite what is needed,
since it includes the axial baryon number symmetry which is not present
in QCD.  This problem is solved [\wilczek ] by restricting ${\cal M}$ to
unitary matrices with positive determinant, instead of general complex
matrices.  Matrices ${\cal M}$ in this restricted class remain in this
restricted class under the transformation \ba\ only if $U$ and $V$ have
equal phases.  Hence, the axial $U(1)$ has indeed been removed.  The
$2 \times 2$ matrices ${\cal M}$ can conveniently be parametrized
using four real parameters $(\sigma ,\vec \pi)$ and the Pauli matrices as
$$
{\cal M}~=~ \sigma + i\vec\pi \cdot \vec\tau~.
\eqn\bc
$$
In fact the order parameter can be written as
a four-component vector $\phi \equiv (\sigma, \vec \pi)$
and the transformations \ba\ are simply $O(4)$ rotations
in internal space.  Hence, the order parameter appropriate for the chiral
phase transition in QCD with two flavors of massless quarks has the
symmetries of the standard $O(4)$ invariant
$N = 4$ Heisenberg magnet.  For smaller number of components,
this sort of model is a much-studied model for the critical
behavior of magnets, with the order parameter representing the
magnetization of a ferromagnet or the staggered magnetization
of an antiferromagnet.

If the phase transition is second order, then it will correspond to
an infrared fixed point of the renormalization group.
We wish to describe those aspects of the critical behaviour which are
universal, that is, those aspects
which are determined by the scaling behaviour of
operators near the infrared fixed point of the renormalization group.
Hence, it is sufficient to retain those degrees of freedom
which develop large correlation lengths at the critical point.
These are just long wavelength
fluctuations of the order parameter, which is small in magnitude near
the critical point and therefore fluctuates at little cost in energy.
Thus the most plausible
starting point for analyzing the critical behavior of a
second-order phase transition in QCD is the Landau-Ginzburg free energy
$$
F = \int d^3 x \Bigl\lbrace {1\over 2}~\partial^i \phi^\alpha
     \partial_i \phi_\alpha
    ~+~ {\mu^2 \over 2} ~\phi^\alpha \phi_\alpha ~+~
    {\lambda \over 4}(\phi^\alpha \phi_\alpha)^2
    ~\Bigr\rbrace ~.
\eqn\bb
$$
Here $\mu^2$ is the temperature-dependent renormalized (mass)$^2$,
which is negative below and positive above the critical point, while
$\lambda$ is the strength of the quartic
couplings and is supposed to be smooth at the transition.
We neglect terms with higher powers of $\phi$ since
$| \phi |$ is small  near the transition. The
symmetry breaking pattern we want is ${\cal M} \propto {\bf 1}$ (equivalently,
$\langle \sigma \rangle \neq 0; \langle \vec \pi \rangle = 0$)
below the
transition which is indeed what we find at the minimum of the
potential for positive $\lambda$.  This model has been studied in depth
for arbitrary $N$ and spatial dimension $d$, and the existence of
an infrared stable fixed point of the renormalization group has been
established [\amitma ].  Hence, it is a model for a second order QCD chiral
phase transition for two massless quarks.

When the free energy \bb\ is written in terms of $\sigma$ and $\vec \pi$
it looks much like the original model of Gell-Mann and Levy [\gm ] with two
changes: there are no nucleon fields and only three (spatial) dimensions.
These two changes reflect an important distinction [\ginsparg ].
We are only proposing
\bb\ as appropriate near the second order phase transition point.  This
is because it is only there that we can appeal to universality -- the
long-wavelength
behaviour of the $\sigma$ and $\vec \pi$ fields is determined by
the infrared fixed point of the renormalization group, and microscopic
considerations are irrelevant to it.
In Euclidean field theory at finite temperature, the integral over $\omega$
of zero temperature field theory is replaced by a sum over Matsubara
frequencies $\omega_n$ given by $2n\pi T$ for bosons and $(2n+1)\pi T$ for
fermions with $n$ an integer.  Hence, one is left with a Euclidean theory
in three spatial dimensions with massless fields from the $n=0$ terms
in the boson sums and massive fields from the rest of the boson sums and
the fermion sums.   Hence, to discuss the massless modes of interest at
the critical point, \bb\ is sufficient.
This means that we do not need to worry
whether to introduce nucleon fields as in [\gm ], or constituent quark
fields as, for example, in [\gocksch ].

We have motivated a very definite hypothesis for the nature of the
phase transition for QCD with two species of massless quarks, namely that
it is in the universality class of the $N=4$ Heisenberg magnet.
This hypothesis has numerous consequences, which are the subject of the
rest of this paper.  To keep the discussion self-contained, in the
remainder of this section we review the predictions for the
static critical exponents described in [\wilczek ].

\section{Critical exponents}

First, we define
the reduced temperature $t~=~(T-T_c)/T_c$.
The exponents $\alpha$,  $\beta$, $\gamma$, $\eta$, and $\nu$
describe the singular
behaviour of the theory with strictly zero quark masses as
$t \rightarrow 0$.
For the specific heat one finds
[6~
$$
C(T) \sim |t|^{-\alpha} + {\rm less~ singular.}
\eqn\da
$$
The behaviour of the order parameter defines $\beta$.
$$
\langle | \phi |\rangle \sim  | t |^{\beta}~~ {\rm for}~ t<0~.
\eqn\dab
$$
$\eta$ and  $\nu$ describe the behaviour of the correlation length $\xi$
where
$$
 G_{\alpha \beta}(x)~ \equiv~
 \langle \phi(x)_\alpha \phi(0)_\beta \rangle
 -\langle \phi_\alpha \rangle \langle \phi_\beta \rangle
 ~\rightarrow~
 \delta_{\alpha \beta} {A \over |x|} \exp (-|x|/\xi) ~~{\rm
at~large~distances.}
\eqn\df
$$
$A$  is independent of $|x|$, but may depend on $t$.
The correlation length
exponent $\nu$ is defined by
$$
\xi \sim |t|^{-\nu }~.
\eqn\de
$$
Above $T_c$, where the correlation lengths are equal in the sigma and pion
channels,
the susceptibility exponent $\gamma$ is defined by
$$
\int d^3 x~G_{\alpha \beta}(x) \sim t^{-\gamma}.
\eqn\deb
$$
We will discuss the behaviour of the susceptibility below the transition
in the following section.
The exponent $\eta$ is defined through the
behaviour of the Fourier transform of the correlation function:
$$
G_{\alpha \beta}(k\rightarrow 0) \sim k^{-2+\eta} ~.
\eqn\dea
$$

The last exponent, $\delta$ is related to the behaviour
of the system in a small magnetic field $H$ which explicitly
breaks the $O(4)$ symmetry.  Let us first show that in a QCD context,
$H$ is proportional to a common quark mass $m_u = m_d \equiv m_q$.
This common mass term may be represented by a $2\times 2$ matrix ${\cal D}$
given by $m_q$ times the identity matrix.
We are now allowed to construct the free energy from invariants involving
both ${\cal D}$ and ${\cal M}$.  The lowest dimension term linear in ${\cal D}$
is just ${\rm tr} {\cal M}^\dagger {\cal D} = m_q \sigma$,
which in magnet language
is simply the coupling of the magnetization to an external field $H \propto
m_q$.  In the presence of an external field, the order parameter is
not zero at $T_c$.  In fact,
$$
\langle | \phi |\rangle (t=0,H\rightarrow 0) \sim H^{1/\delta}~.
\eqn\dec
$$

The six critical exponents defined above are related by four scaling
relations [\amitma ].  These are
$$
\eqalign{
\alpha~&=~2-d\nu \cr
\beta~&=~{\nu \over 2} ( d - 2 + \eta ) \cr
\gamma~&=~(2-\eta )\nu \cr
\delta~&=~{d + 2 - \eta \over d - 2 +\eta}~. \cr }
\eqn\db
$$
We therefore need values for $\eta$ and $\nu$ for the four component
magnet in $d=3$. These were obtained in the remarkable work of
Baker, Meiron and Nickel[\bmn ], who carried the perturbation theory to
seven-loop order, and used information about the behaviour of
asymptotically large orders, and conformal mapping and Pad\'e
approximant techniques to obtain
$$
\eqalign{
\eta ~&=~~ .03 \pm .01 \cr
\nu  ~&=~~ .73 \pm .02 ~.  \cr }
\eqn\dc
$$
Using \db\ , the remaining exponents are
$\alpha = -0.19 \pm .06$, $\beta = 0.38 \pm .01$,
$\gamma = 1.44 \pm .04$ and $\delta = 4.82 \pm .05$.  Since $\alpha$
is negative there is a cusp in the specific heat at $T_c$, rather than a
divergence.

Very different critical exponents are proposed in [\kocic ].
These authors thus implicitly claim that a
hitherto unknown fixed point theory
with the symmetries of the $N=4$ isotropic Heisenberg model exists,
and governs the QCD phase transition.

\chapter{The Equation of State and the Pion and Sigma Masses}

The expressions which define $\beta$, $\gamma$ and $\delta$
are actually special cases of a more general relationship between
the magnetization and the magnetic field called the critical
equation of state.  The equation of state has been calculated to
order $\epsilon^2$ by Br\'ezin, Wallace and Wilson [\bww ].  Their
result is reproduced in the appendix.  In this section, we will
use the equation of state to determine the behaviour of the masses
of the pion and sigma masses near the critical point.

First, we must define what we mean by the ``mass'' of the pion
and sigma.   We could choose either to define the mass as an inverse
correlation length or as an inverse susceptibility.  We choose the
latter, which is conventional in the condensed matter literature.
Specifically, we define
$$
m_{\sigma}^{-2} = \int d^3 x G_{0 0}
\eqn\ea
$$
and
$$
m_{\pi}^{-2} \delta_{ij} = \int d^3 x G_{ij}
\eqn\eb
$$
where $\phi_0 = \sigma$ and $\phi_i = \pi_i, i=1,2,3$.  At any given $t$
and $H$, \df\ implies that whether one defines the mass as the inverse
correlation length or as the inverse susceptibility is academic.  However,
since $A$ in \df\ depends on $t$ and $H$, the two different choices lead
to different scaling behaviours for masses as functions of $t$ and $H$.
We shall see that with the conventional choice \ea\ and \eb\ ,
the masses can be extracted conveniently from the equation of state.
It is worth noting that the masses we have defined are related only
to the behaviour of spatial correlation functions in the static
(equilibrium) theory.  They carry no dynamical information.  Also,
we will only be able to make universal statements about how the masses
scale at the transition.  Normalizing the magnitudes of the masses ({\it i.e.}
relating them to the zero temperature masses) will require using some
specific model, and hence will not be universal.

The equation of state gives the magnetization as a function of $t$ and
$H$.  For the rest of this paper, we will write the order parameter
as $M$, for magnetization, keeping in mind that $M = \langle \sigma \rangle
= \langle |\phi| \rangle $.  In order to define the equation of
state, we first define a shifted field $\tilde \sigma = \sigma - \langle
\sigma \rangle = \sigma - M$.  Then the equation of state is simply
the relation
$$
\langle \tilde \sigma \rangle = 0.
\eqn\ec
$$
This relation has been expanded to order $\epsilon^2$ by Br\'ezin, Wallace
and Wilson [\bww ].  The result
can be expressed conveniently in terms of
the variables $y \equiv H/M^\delta$ and $x\equiv t/M^{1/\beta}$ as
$$
y = f(x)
\eqn\ed
$$
where the function $f(x)$ was calculated to order $\epsilon^2$ in [\bww ],
and is given in the appendix.  The units in which $H$ and $M$ are
measured are chosen so that $f(0) = 1$ corresponds to $t=0$ and
$f(-1)=0$ corresponds to the $t<0,~H=0$ coexistence curve.
Knowing $f(x)$, we can calculate the value of the order parameter $M$
for a given $H$ and $t$ using \ed\ .
The behaviour of the order parameter is illustrated in Figure 1.
This figure and the other ones in this section should be viewed as
illustrations
of qualitative behaviour rather than quantitative predictions
because they are based on setting $\epsilon =1$
in the $O(\epsilon ^2)$ expression for $f(x)$. The values for the
critical exponents themselves which we quoted in the previous section are
quantitative predictions, complete with error estimates, because they are
based on the much more elaborate analysis of Baker {\it et al.} [\bmn ]

{}From the equation of state, we can deduce the behaviour of $m_\pi$
and $m_\sigma$ at non-zero (but small) $t$ and $H$.
The masses are given by
$$
m_\sigma^2 = { \partial H \over \partial M}
\eqn\ee
$$
and
$$
m_\pi^2 = {H \over M}.
\eqn\ef
$$
The first relation follows directly from the definition \ea\ , and
the second follows from \eb\ and
from assuming that $\vec M \parallel \vec H$, so that a small change $\delta H
\perp H$ gives a small change $\delta M \perp M$ with $\delta M / \delta
H = M/H$.
Using the equation of state, we can rewrite \ee\ and \ef\ as
$$
m_\pi^2 = M^{\delta - 1} f(x)
\eqn\efb
$$
and
$$
m_\sigma^2 = M^{\delta - 1} \left( \delta f(x) - {x \over \beta} f'(x)
\right)~.
\eqn\efc
$$
Hence, as suggested in [\kocic ], $\delta$ can be determined by measuring the
ratio $m_\sigma^2 / m_\pi^2 $ at $t=0$.  In general, from $f(x)$ we can find
the pion and sigma masses for any $t$ and $H$.

There are two interesting limits which we will consider
explicitly.  First, for $t>0$ and $H\rightarrow 0$ which corresponds
to $x\rightarrow \infty $, we should find the full $O(4)$ symmetry,
and hence should find that the pion and sigma masses are identical.
For $x\rightarrow \infty $, the function $f(x)$ from the appendix
behaves as $f(x)=c x^\gamma $.  Here, the constant $c$ and the
exponent $\gamma$ are given to $O(\epsilon)$ and $O(\epsilon^2)$
respectively in (A9) and (A10) .
Applying
\efb\ and \efc\ , we find that
$$
m_\sigma^2 = m_\pi^2 = c t^\gamma~~{\rm for~}x\rightarrow \infty ~,
\eqn\eg
$$
consistent with the symmetry.

We can also consider the limiting case of approaching the coexistence curve.
This means taking $t<0$ and $H\rightarrow 0$, which implies
$x\rightarrow -1$.  In this limit, $M$ tends to a nonzero constant,
and so from \ef\ , we obtain $m_\pi^2 \propto H$, a familiar result for
Goldstone bosons.  The behaviour of the pion mass is illustrated in Figure 2a.
The result \ef\ may look peculiar to a particle physicist who is more familiar
with the zero temperature result
$$
m_\pi^2 = {2 m_q \langle \bar q q \rangle \over f_\pi^2}~.
\eqn\ega
$$
Before considering the sigma mass, we therefore pause here to explain how
\ega\ and \ef\ are related. We have seen that $m_q \sim H$ and that
the order parameter $\langle \bar q q \rangle \sim \langle \sigma \rangle
\sim M$.
At zero temperature,
$f_\pi$ is defined in terms of the axial current by the relation
$$
\langle 0 \mid A_\mu^\alpha (0) \mid \pi^\beta (q) \rangle =  i f_\pi q_\mu
\delta^{\alpha\beta}~.
\eqn\egb
$$
In the zero temperature linear sigma model, the axial current is given by
$$
A_\mu^\alpha (x) = \sigma (x) \partial_\mu \pi^\alpha (x) ~-~ \pi^\alpha (x)
\partial_\mu \sigma (x)~,
\eqn\egc
$$
which means that $f_\pi$ defined in \ega\ is simply
$$
f_\pi = \langle 0 \mid \sigma \mid 0 \rangle ~.
\eqn\egd
$$
This result suggests that we make the identification $f_\pi \sim M$, which
does indeed make \ef\ and \ega\ equivalent.  However, it is important
to remember that
using the linear sigma model at zero temperature can not be justified by
a universality argument in the way that using it near $T=T_c$ can.
Hence the argument of this paragraph is {\it not} a derivation of \ef\
from the zero temperature result \ega\ .  \ef\ is valid near $T=T_c$ while
\ega\ is valid at $T=0$.  Also, $m_\pi$ in \ega\ is a mass in a $3+1$
dimensional Lorentz invariant theory,
while $m_\pi$ in \ef\ is an inverse susceptibility in a 3 dimensional
theory.
We have simply shown that a reader familiar
with one expression should not be surprised by the other.

The behaviour of the sigma mass at the coexistence
curve is trickier to obtain than that of the pion mass.
First, we note that in mean field theory ($\epsilon = 0$) the equation
of state is simply $y = f(x) = 1+x$, and $m_\sigma$
is easily evaluated using \efc\ .  For $H\rightarrow 0$ at fixed $t<0$
the result is
$$
m_\sigma^2 = \Bigl( {\delta \over |t|^\beta } \Bigr) H
+ {|t|^{\beta (\delta -1 )} \over \beta } ~.
\eqn\ege
$$
Hence, in mean field theory $m_\sigma^2$ decreases with $H$ to a non-zero
value at $H=0$.
However, for $d<4$ when
fluctuations are important, the result is quite different.  In words,
fluctuations of the massless pions produce new infrared singularities
in the longitudinal susceptibility, or, equivalently,
make the sigma massless. Now, let us see how this result can
be obtained from the equation of state [\bw ,\wz ].
In the limit $H\rightarrow 0$, $f(x) \sim H$ while $f'(x)$, we will see,
tends to zero more slowly.  Hence, the second term in \efc\ is dominant and
gives
$$
{ \beta m_\sigma^2 \over M^{\delta -1}} \rightarrow f'(x)
{}~{\rm for}~x\rightarrow -1~.
\eqn\eh
$$
The difficulty is that from the expression (A11) for $f(x)$ valid for
$x\rightarrow -1$,
we notice that $f'(x)$ contains divergent terms like $\epsilon \log(x+1)$,
 $\epsilon^2 \log ^2(x+1)$ and $\epsilon^2 \log(x+1)$. These terms do {\it not}
exponentiate to $f'(x) \sim (x+1)^p$, contrary to the claims of [\kocic ].
After some algebra [\wz ], one
finds the result
$$
\left( {\beta m_\sigma^2 \over M^{\delta -1}} \right) ^{-1}
\rightarrow c_1 + c_2 y^{-\epsilon /2} ~.
\eqn\ei
$$
Both the terms on the right side of \ei\ must be kept because they
differ in their exponents only by order $\epsilon$.  Also for this
reason,
the constants $c_1$ and $c_2$ given in the appendix are known only
to order $\epsilon$ even though $f(x)$ is known to order $\epsilon ^2$.

Qualitatively, as $H$ is lowered at fixed
$t<0$, at first the $c_1$ term dominates and $m_\sigma^2$ appears to be
decreasing toward a non-zero value at $H=0$ as in the mean field result.
Then, the $c_2$ term takes over and one finds that in fact the sigma
mass goes to zero like $m_\sigma^2 \propto H^{\epsilon/2}$.
The behaviour of the sigma meson mass is illustrated
in Figures 2b and 3.
In future lattice simulations, as $m_q$ is lowered toward zero, this behaviour
should be observed.
This result is an example of the power of the renormalization group
techniques in obtaining universal results.  If we had chosen a specific
model, say that of Gocksch [\gocksch ], or the Nambu-Jona-Lasinio
model of Hatsuda and Kunihiro [\hk ], we would have been able to
calculate non-universal quantities far from $T_c$, but would basically
have been limited to using mean field theory, as those authors do.
Then, we would have reached the incorrect conclusion that the sigma
has a non-zero mass in the chiral limit below $T_c$.  Here, by restricting
ourselves to calculating universal quantities, we are limited to
the region near the critical point, but we can be confident in our
results regardless of which specific model is correct and can
include the effect of fluctuations.

\chapter{$\rho$ and $a1$ Correlation Functions}

To this point, we have discussed the correlation functions in the pion
and sigma channels only.  It is certainly possible to construct other
spatial correlation functions.  The next-simplest are those
associated with the $\rho$ (Lorentz vector, isospin vector) and
$A_1$ (Lorentz vector, isospin axial vector) mesons.  In Gocksch's
model [\gocksch ], these have correlation lengths given
simply by $2 m_Q (T)$ where $m_Q = \pi T + g\langle \sigma \rangle$
is the mass of the constituent quarks in his model. Since these constituent
quarks are fermions, they have a Matsubara mass of $\pi T$. $g$ is a
coupling constant.  In any model, there is bound to be a model dependent
contribution to these correlation lengths which is smooth at $T_c$.
However, the pions and sigma also contribute to the $\rho$ and $A_1$
correlation
lengths since operators with the appropriate symmetries can be constructed
from the pion and sigma operators.  In particular,
$$
\rho_\alpha^i = \epsilon_{\alpha \beta \gamma}\pi_\beta \nabla^i
\pi_\gamma
\eqn\fa
$$
and
$$
(A_1)_\alpha^i = \sigma \nabla^i \pi_\alpha - \pi_\alpha \nabla^i
\sigma ~~.
\eqn\fb
$$
Hence, the pions and sigma can make a universal non-analytic contribution
to the $\rho$ and $A_1$ correlation lengths at the critical point.
In this section, we calculate this contribution for $t>0$ and $H=0$.

Although we will try to define all the quantities we use, for those interested
in further details we note that we are following the notation and conventions
of Amit [\amitma ].  In the region $t>0$ and $H=0$ the symmetry is unbroken
and hence there is no distinction between the $\sigma$ and $\pi$ operators
which
we will call $\phi^\alpha$ and the $\rho$ and $A_1$ operators which
we combine into ${\cal O} _i^{\alpha\beta} \equiv
\epsilon^{\alpha\beta\gamma\delta} \phi^\gamma \nabla _i \phi^\delta$.
We are interested in the scaling behaviour of the correlation function
$\langle {\cal O} (x) {\cal O} (0) \rangle \equiv
\Gamma^{(0,0,2)}$. In general, by $\Gamma^{(m,n,p)}$ we mean the $m$-point
vertex function with $n$ insertions of $\phi^2$ and $p$ insertions
of $\cal O$.  The scaling behaviour
of vertex functions involving a composite operator like $\cal O$ is
determined by $\eta$, $\nu$, and the anomalous dimension of the operator.
Here we are fortunate because
the operators ${\cal O}_i$
are the ``conserved'' currents of the chiral symmetry, where
in the 3 dimensional theory this means $\nabla^i {\cal O}_i = 0$.
The Ward identities arising from chiral symmetry imply that the $\cal O$
are not subject to renormalization [\pokorski ], or, equivalently, that their
anomalous dimensions are zero.

For $H=0$, the vertex function $\Gamma^{(0,0,2)}$ is a function
of the external momentum $k$, the reduced temperature $t$, the
renormalization point $\kappa$, and the $\phi^4$ coupling constant
$\lambda$.  At the fixed point of the renormalization group, $\lambda$ is a
constant and the vertex function satisfies the renormalization
group equation [\amitma ]
$$
\Biggl \lbrace \kappa {\partial \over \partial \kappa} - \Bigl (
{1 \over \nu} - 2 \Bigr ) t {\partial \over \partial t }
\Biggr \rbrace \Gamma^{(0,0,2)}(k,t,\kappa)~=~B~.
\eqn\fc
$$
The right hand side of the equation is non-zero because although
$\Gamma^{(0,0,2)}$ is not subject to multiplicative renormalization,
it is additively renormalized.  However, $B$ does not depend on either
$k$ or $t$ [\amitma ], and therefore does not contribute to the
non-analytic behaviour at the critical point. Therefore, in what follows
we will drop $B$.
As a consequence of \fc\ , the vertex function has the form
$$
\Gamma^{(0,0,2)}(k,t,\kappa) = F\Bigl ( ~\kappa^{1/\theta }
{}~,~k~\Bigr ) ~,
\eqn\fd
$$
where $\theta = {1 \over \nu} -2$.
By dimensional analysis, we know that
$$
\Gamma^{(0,0,2)}(k,t,\kappa) = b^{d-2} \Gamma^{(0,0,2)}\Bigl (
{k \over b} , {t \over b^2 }, {\kappa \over b} \Bigr )~.
\eqn\fe
$$
This means that \fd\ becomes
$$
\Gamma^{(0,0,2)} = b^{d-2} F\Bigl ( ~{\kappa \over b} \bigl (
{t \over b^2 } \bigr )^{1/\theta }~,~{k \over b}~\Bigr )~.
\eqn\ff
$$
To this point, $b$ has been arbitrary.  Therefore, we can choose it
strategically.  With the choice $b=\kappa (t/\kappa^2 )^\nu$,
we find
$$
\Gamma^{(0,0,2)} \sim t^{\nu (d-2)} g(kt^{-\nu})
\eqn\fg
$$
where the function $g(x)$ satisfies $g(0)={\rm constant}$.
If, on the other hand, we choose $b=k$, the result is
$$
\Gamma^{(0,0,2)} \sim k^{d-2} \tilde g (kt^{-\nu})
\eqn\fh
$$
where the function $\tilde g(x)$ tends to a constant for $x\rightarrow \infty$.
\fg\ , which gives the behaviour of $\Gamma^{(0,0,2)}$ as a function
of $t$ for $k=0$, and \fh\ , which gives the behaviour for $t=0$
as a function of $k$, describe the non-analytic part of the correlation
function.  As we mentioned earlier, there will also be a model dependent
but analytic mass term for the $\rho$ and $A_1$. Except near the critical
point,
this smooth term
is presumably larger
than the non-analytic term whose effects we have calculated.
In order to observe \fg\ and \fh\ , future lattice simulations will have
to get close enough to the critical point that the non-analytic term
dominates the analytic term.

\chapter{The Influence of the Strange Quark}

To this point in this paper, we have described a world with two
massless quarks, and hence we have implicitly been taking the
strange quark mass to be infinite.  If the strange quark is massless,
then Pisarski and Wilczek showed [\wilczek ] that the chiral phase transition
is first order.  Hence, as the strange quark mass is reduced from
infinite to zero, at some point the phase transition must change from
second order to first order.  This point is called a tricritical point.
There is numerical evidence [\columbia ]
that when the strange quark has its physical mass, the transition
is second order.  Hence, we devote most of this paper to analyzing
the second order phase transition.  However, in a lattice simulation,
the strange quark mass could be tuned to just the right value to reach
the tricritical point.  In this section, we discuss the critical
exponents that would be observed in such a simulation.

Let us consider the effect of adding a massive but not infinitely
massive strange quark to the two flavor theory.  This will not introduce
any new fields which become massless at $T_c$, and so the arguments
leading to the free energy \bb\ are still valid.  The only effect of the
strange quark, then, is to renormalize the couplings.  Renormalizing
$\mu^2$ simply shifts $T_c$, as does renormalizing $\lambda$ unless
$\lambda$ becomes negative.  In that case, one can no longer truncate
the Landau-Ginzburg free energy at fourth order.
After adding a sixth order term, the free energy becomes
$$
F = \int d^3 x \Bigl\lbrace {1\over 2}(\nabla \phi)^2 + {\mu^2 \over  2}\phi^2
+ {\lambda \over 4}(\phi^2)^2 + {\kappa \over 6}(\phi^2)^3 - H\sigma
\Bigr\rbrace ~.
\eqn\ga
$$
While for positive $\lambda$, $\phi^2$ increases continuously from zero
as $\mu^2$ goes through zero, for negative $\lambda$, $\phi^2$
jumps discontinuously from zero to $|\lambda |/(2\kappa)$ when
$\mu^2$ goes through $\lambda^2 /(4\kappa)$.  Hence, the phase transition
has become first order.  Thus at the value of $m_s$ where $\lambda = 0$,
the phase transition changes continuously from second order to first order.

The singularities of thermodynamic functions near tricritical points, like
the singularities near ordinary critical points, are universal.  Hence,
it is natural to propose [\wilczek ] that QCD with two
massless flavors of quarks and with $T$ near $T_c$ and $m_s$ near its
tricritical value is in the universality class of the $\phi^6$ Landau-Ginzburg
model \ga\ .  This model has been studied extensively [\lawrie ].
Because the $\phi^6$ interaction is strictly renormalizable in three
dimensions,
this model is much simpler to analyze than the $\phi^4$ model of the
ordinary critical point.  No $\epsilon$ expansion is necessary, and
the critical exponents all take their mean field values.  There are calculable
logarithmic corrections to the scaling behaviour of thermodynamic
functions [\lawrie ], but we will limit ourselves here to determining
the mean field tricritical exponents.

In mean field theory, the correlation function in momentum space is simply
$G_{\alpha \beta}(k) = \delta_{\alpha \beta} (k^2 + \mu^2)^{-1}$.
Since $\mu^2 \sim t$, this gives the exponents $\eta =0$, $\gamma =1$
and $\nu = 1/2$.  To calculate $\alpha$ and $\beta$, we minimize
$F$ for $H=\lambda=\nabla \phi =0$, and find $\alpha = 1/2$ and
$\beta = 1/4$.
To calculate $\delta$, we minimize $F$ for $t=\lambda =\nabla \phi =0$
and find $\delta = 5$.

The result for the specific heat exponent $\alpha$
is particularly interesting, since it means that
the specific heat diverges at the tricritical point, unlike at the
ordinary critical point.
This means that whereas for $m_s$ large enough that the transition
is second order the specific heat $C(T)$ has a cusp but is finite
at $T=T_c$, as $m_s$ is lowered to the tricritical value $C(T_c)$
should increase since at the tricritical point it diverges.  This behaviour
should be seen in future lattice simulations.

Finally, at a tricritical point there is one more
relevant operator than at a critical point, since two physical quantities
($t$ and $m_s$) must be tuned to reach a tricritical point.  Hence,
a new exponent $\phi_t$, the crossover exponent, is required.
For $\lambda \neq 0$, tricritical behaviour
will be seen only for $|t|>t^*$, while for $|t|<t^*$, either ordinary
critical behaviour or first order behaviour (depending on the sign
of $\lambda$) results.  $t^*$ depends on $\lambda$ according to
$$
t^* \sim \lambda^{1/\phi_t}
\eqn\gb
$$
The mean field value of $\phi_t$ is obtained by minimizing the free
energy $F$ for $H=\nabla \phi =0$, and is $\phi_t = 1/2$.
These mean field tricritical exponents, $\alpha = 1/2$, $\beta = 1/4$,
$\gamma = 1$, $\delta = 5$, $\eta = 0$, $\nu = 1/2$, and $\phi_t = 1/2$
would describe the real
world if $m_s$ were smaller than it is, and will describe
future lattice simulations with $m_s$ chosen appropriately.

\chapter{Dynamic Critical Phenomena}

To this point, we have discussed the static critical phenomena
appropriate to the equilibrium properties of the
QCD plasma near its phase transition.  To discuss dynamical phenomena, we
need equations of motion.  In zero temperature scalar field theory, Lorentz
invariance requires that these equations have no first order time
derivatives.  In the finite temperature theory appropriate for discussing
critical behaviour, however, there is no Lorentz invariance, and hence we
should expect first order time derivatives.
In analogy with the notion of static universality,
one finds that the dynamics of the long wavelength modes groups theories
into dynamic universality classes containing theories described by the
same equations of motion.
To specify the dynamic universality class, one needs to specify more
than the dimension of space and the number of components of the
order parameter.  Hence, there
are usually several different dynamic universality classes
which are all in the same static universality class.
In particular, it is necessary to specify whether or not the order parameter is
conserved, and which other quantities are conserved.
Thus a ferromagnet will have
very different dynamical behaviour from an antiferromagnet with the
same number of components, even though the static universality class
is the same.  The ferromagnet will
have a much more difficult time thermalizing
long-wavelength fluctuations, since in the $k\rightarrow 0$ limit
they are rigorously stable.

{}From this point of view, two flavor QCD behaves as an {\it antiferromagnet}.
Its order parameter, the expectation value of a
scalar or pseudoscalar quark bilinear, is
not a conserved quantity.  (It is
quantities of the form $\bar q \gamma_0 q$, not $\bar q q$, that are
conserved).
This means that a model similar to model G
of Halperin and Hohenberg [\hh ], is appropriate.  Model G is formulated
for a three component order parameter.  We show below how to formulate it
for the $N=4$ order parameter appropriate for QCD.
One main result
of the theory concerns the rate of critical slowing down near the
transition.  Just as the correlation length in space diverges at the
critical point, so does the correlation time for dynamics.  Its scaling
property is conventionally written in terms of a critical exponent $z$,
such that the correlation time scales as $\xi^z$, where $\xi$ is
the correlation length.  For the model in question, we show below that
the exponent $z$ governing critical slowing down is predicted to be
$d/2~=~ 3/2 $.

Let us now consider how model G of [\hh ] must be modified to deal with
the four component order parameter of interest.  As we discussed above,
it is important to find the conserved quantities.  First, there is the
energy.  However, it is shown in [\hhm ] that if $\alpha$, the specific
heat exponent, is negative then the dynamics of the order parameter
is not affected by the presence of a conserved energy.
However, the six generators of $O(4)$ rotations are associated with
six conserved quantities which can be written as an antisymmetric
tensor
$$
J_{\alpha \beta } = \int d^3 x j_{\alpha \beta}~.
\eqn\ha
$$
At the level of the unrenormalized Lagrangian, $j_{\alpha}$ can be expressed
in terms of the order parameter as
$$
j_{\alpha \beta } = \epsilon_{\alpha \beta \gamma \delta}\phi_{\alpha}
{\partial \over \partial t}\phi_{\beta} ~,
\eqn\hb
$$
but this relation need not hold upon renormalization.  (Note that
since we are interested in the critical phenomena, by renormalization
we mean renormalization towards the infrared.)
In model G, the symmetry is $O(3)$, so there are 3 conserved quantities
instead of 6.  The 3 conserved quantities are the magnetization, while
the nonconserved order parameter is the staggered magnetization of the
antiferromagnet.
The equations of motion for $j$ and $\phi$ contain two types of
terms.  There are dissipative terms which damp the system toward
the equilibrium configuration and so called mode-mode coupling
terms.  The latter reflect the Poisson bracket relations among
the fields.  In our case, these are
$$
\Bigl[ \phi_\alpha , J_{\beta \gamma} \Bigr] = \epsilon_{\alpha
\beta \gamma \delta} \phi_\delta
\eqn\hc
$$
and
$$
\Bigl[ J_{\alpha \beta} , J_{\gamma \delta} \Bigr] = \delta_{\beta \gamma}
J_{\alpha \delta} - \delta_{\alpha \gamma} J_{\beta \delta} -
\delta_{\beta \delta} J_{\alpha \gamma} + \delta_{\alpha \delta}
J_{\beta \gamma} ~.
\eqn\hd
$$
The equations of motion are
$$
{\partial \phi_\alpha \over \partial t} = -\Gamma {\delta F \over \delta
\phi_\alpha } + g \Bigl[ \phi_\alpha , j_{\beta \gamma } \Bigr] {\delta F
\over \delta j_{\beta \gamma}} + \theta
\eqn\he
$$
and
$$
{\partial j_{\alpha \beta} \over \partial t} = \gamma \nabla^2 {\delta F
\over \delta j_{\alpha \beta}} + g \Bigl[ j_{\alpha \beta} , \phi_{\gamma}
\Bigr] {\delta F \over \delta \phi_\gamma } + g \Bigl[ j_{\alpha \beta} ,
j_{\gamma \delta} \Bigr] {\delta F \over \delta j_{\gamma \delta} } + \zeta
{}~,
\eqn\hf
$$
where the free energy $F$ is given by
$$
F = \int d^3 x \Bigl\lbrace {1\over 2}~\partial^i \phi^\alpha
     \partial_i \phi_\alpha
    ~+~ {\mu^2 \over 2} ~\phi^\alpha \phi_\alpha ~+~
    {\lambda \over 4}(\phi^\alpha \phi_\alpha)^2 ~+~ { 1 \over 2 \chi }
    j_{\alpha \beta} j^{\alpha \beta} ~-~ H\sigma
    ~\Bigr\rbrace ~.
\eqn\hg
$$
$\theta$ and $\zeta$ are Langevin noise terms.
The difference between the dissipative terms in the two equations of motion
reflects the fact that $j$ is conserved, and hence a spatially constant
$j$ can not dissipate.
Note that $j$ appears in the free energy only as $j^2$.  This is
because any higher order terms like $j^2 \phi^2$ or those involving
$\nabla j$ are irrelevant.  Because of the form of $F$, it
turns out that the term in the equation of motion \hf\ involving
\hd\ is zero.
The equations of motion given above determine the universal dynamics of the
long wavelength modes of interest near the critical point.

With equations of motion in hand, Halperin {\it et al.} [\hh ] go on to
formulate dynamic renormalization group transformations.
These are obtained by starting with the theory defined with an ultraviolet
cut-off $\Lambda$, integrating out modes in the momentum shell
between $b^{-1} \Lambda$ and $\Lambda$ and then rescaling according to
$$
\eqalign{
x ~&\rightarrow x' ~=~ b^{-1} x \cr
\phi ~& \rightarrow \phi ' ~=~ b^{(d-2+\eta)/2} \phi \cr
j ~&\rightarrow j' ~=~ b^c j \cr
t ~&\rightarrow t' ~=~ b^{-z} t ~. \cr }
\eqn\hh
$$
After one such transformation, one obtains a new free energy and new equations
of motion.  If one works in the $\epsilon$ expansion, these are related
to the former $F$ and equations of motion simply by a transformation
of the parameters $\mu$, $\lambda$, $\Gamma$, $\gamma$, $g$, and $\chi$.
Repeated application of the renormalization group transformation therefore
leads to recursion relations for these parameters.
The next step is to find the fixed point of the transformation.  Just as
in the static case the fixed point condition fixes $\eta$, here it
fixes $\eta$, $c$, and $z$.
Fortunately, to obtain $c$ and $z$ we only need the particularly simple
recursion relations for $\chi$ and $g$.  Because $j$ only appears quadratically
in the free energy, the recursion relation for $\chi$ is
$$
\chi^{-1} \rightarrow b^{d-2c} \chi^{-1}~,
\eqn\hi
$$
and this is valid to all orders in $\epsilon$.
The terms in the equations of motion proportional to $g$ are consequences
of the $O(4)$ symmetry, and as a result the Ward identities enforce the
recursion relation
$$
g \rightarrow b^{z-d+c} g
\eqn\hj
$$
to all orders in $\epsilon$.
{}From \hi\ and \hj\ it is clear that at a fixed point one must have
$c=d/2$ and, as advertised earlier, $z=d/2$.

\chapter{Implications on the Lattice and in the Real World}

Finite temperature lattice QCD simulations are ideally suited to
testing many of the predictions made in this paper.
The static correlation functions of the three dimensional theory are
natural objects to consider in finite temperature (Euclidean)
simulations.  Also, it is much easier to vary parameters like the
temperature and the bare quark mass in a lattice simulation than
in a real experiment.
Hence, it should be
possible to measure the static critical exponents of section 2,
the equation of
state and the scaling behaviour of the pion and sigma masses of section 3,
the behaviour of the $\rho$ and $A_1$ correlation functions of section 4,
and the static tricritical exponents of section 5.

Present simulations [\columbia ,\bernard ] provide strong evidence that the
QCD phase transition is second order, and that the order parameter
is ${\cal M}$ [\wilczek ].  However, there are several reasons why
present simulations can not yet be used to test the more detailed
predictions of this paper.  The fundamental reason is that in all
simulations to date, the bare quark mass has been so large that
correlation lengths do
not get very long at $T_c$.  For example, in the work of Bernard
{\it et al.} [\bernard ], the correlation length in the pion channel at
$T_c$ is only about 2.5 lattice lengths.  Hence, in order
to study the behaviour nearer to the critical point and to measure
universal properties, we must wait for simulations with smaller
quark masses.  The fundamental problem is that long correlation lengths
are necessarily accompanied by numerical critical slowing down, and this
makes simulations challenging.

Another hurdle to be overcome before
lattice simulations can measure the critical properties of the QCD
phase transition is that any lattice implementation of fermion
fields only exhibits the full chiral symmetry in the continuum limit.
If Wilson fermions are used, there is no chiral symmetry at all on the
lattice.  If Kogut-Susskind fermions are used, four flavors of fermions
are required, but there is a continuous $U(1)\times U(1)$ chiral symmetry
on the lattice.  This should give a phase transition in the
universality class of the $N=2$ magnet, and has been discussed in [\boyd ].
In order to study two flavors of fermions, one takes the square root
of the fermion determinant in the lattice action.  It is not at all
clear what this does to the lattice chiral symmetries.  Hence, for
Kogut-Susskind as for Wilson fermions, before we are able to test
our predictions for the critical phenomena we need finer lattices
so that extrapolation to the continuum limit can be done.
A necessary condition for doing this is that the results for Kogut-Susskind
and Wilson fermions agree when so extrapolated.  This condition has not
yet been met [\wilsonquarks ].

Hence, while we cannot test our detailed predictions against current
lattice simulations, we are confident that in the future with finer
lattices and longer correlation lengths, simulations will be able to
measure critical exponents, correlation functions, and the equation
of state, and verify our results.

Let us now turn to real experiments, as opposed to
those on the lattice. Here, we are not free to dial the
bare quark mass.  We will see that this is unfortunate, particularly
in our discussion of heavy ion collisions.  First, however, let us dispose
of two other possible arenas for testing our results.  The QCD phase transition
certainly occurred in the early universe.  Indeed, much work has been
done on possible observable effects of this transition, if it is first
order.  Unfortunately, for physical values of the quark masses, we
have seen that the transition is second order.  We can think of no observable
consequences of a second order QCD phase transition in the early universe.
In [\wilczek ], it was noted that since certain antiferromagnets including
dysprosium have order parameters in the $N=4$ universality class [\bak ],
experiments on the phase transition in these materials would
help us understand the QCD transition.  Alas, in these magnets there is
a quartic operator like $(\pi_1^2 + \pi_2^2 )(\pi_3^2 + \sigma^2 )$
which is allowed by the microscopic hamiltonian
of the magnets [\barak ], but not by that of QCD.  This operator makes
the symmetric Heisenberg fixed point infrared unstable, and either
makes the phase transition first order or makes it second order but governed
by an anisotropic fixed point.  Hence, neither cosmology nor dysprosium
are suitable arenas for learning about the QCD phase transition.

Now, we turn to relativistic heavy ion collisions.  In Bjorken's [\bjorken ]
picture of such a collision, a volume of hot plasma forms and
quickly reaches thermal equilibrium.
In the center of mass frame, the incident nuclei are both Lorentz contracted
into pancake shapes.  They pass through each other, and leave behind
a region of hot vacuum.  In translation, this means that the baryon number
of the incident nuclei ends up in that part of the plasma heading
approximately down the beam pipes, and the central rapidity region consists
of plasma with approximately zero baryon number.  In the remainder of this
paper, we will
attempt to use what we have learned about the critical
phenomena associated with the chiral phase transition to study
the behaviour of the plasma in the central rapidity region as it expands and
cools through $T=T_c$ and eventually hadronizes and becomes pions which
fly off and are detected.

The defining characteristic of a second order phase transition is
the divergence of correlation lengths.  How could this feature be
observed here?  Large volumes of space with the order parameter
correlated and pointing in a direction different than the true
vacuum ({\it i.e.} sigma) direction will become regions in which the
order parameter oscillates coherently about the sigma direction.
After hadronization, correlated volumes will
turn into regions of space where
the ratio of the number of charged pions to neutral pions has some fixed value.
Since, in the standard scenario of Bjorken [\bjorken ], different positions
in the plasma along the beam direction become different longitudinal
momenta (actually, different rapidities) as the plasma expands, one would
hope that a signal of a second order phase transition would be fluctuations
in the ratio of charged to neutral pions as a function of rapidity.
Coherent evolution of a classical pion field has been considered before
by Anselm and Ryskin [\anselm ] and by Blaizot and Krzywicki [\blaizot ].
However, these authors considered neither an equilibrium second order
phase transition as we do in this section, nor a quench as we will in the
next section.

A prerequisite for the fluctuations discussed above to be observable is
that the correlation length must get long compared, say, to $T_c$.
To determine whether this does indeed happen, we must leave our
universality safety net behind, since neither $T_c$ nor the magnitudes
of correlation lengths are universal.
{}From Figure 2a, it is immediately obvious that we have a problem.
The longest correlation length is in the pion channel, and the
pion mass is increasing with temperature.  This suggests that the
pion mass at $T_c$ is larger than $m_\pi (T=0) ~=~135~{\rm MeV}$.
This is consistent with the fact that in many models the pion mass increases
from its zero temperature value as the temperature is increased from
zero.  This result has been obtained in chiral perturbation theory
using the nonlinear sigma model [\leutwyler ] and also for the linear sigma
model [\loewe ].  Hence, it seems clear that the longest correlation length
at $T_c$ will be shorter than $(135 ~{\rm MeV})^{-1}$.  This is to be
compared to $T_c$ itself, which for the case of two massless quarks
is around 140 MeV [\bernard ].
Hence, even though the quark masses are indeed
small ($\sim 10 ~{\rm MeV}$), the magnetic field $H$ proportional to
$m_q$ is large enough to prevent any correlation lengths from reaching
interesting values.

There is an appropriate quantitative criterion to determine whether
a near equilibrium second order phase transition
leads to dramatic effects.  One compares the energy in a correlation
volume just below $T_c$ with the zero temperature pion mass to
determine whether or not the correlated volume can  become a large
number of pions.  Using current lattice simulations [\bernard ],
we can make a crude attempt
at this comparison.  The sum of the energy and pressure in a correlation volume
is about $1/4$ the zero temperature $\rho$ mass.  Taken literally, this means
that each correlation volume becomes only one or two pions in the detector.
As we have mentioned, current simulations are subject to many caveats, and
so this estimate should not be taken literally.  Nevertheless, it seems clear
that the physical value of $m_q$ is large enough that in an equilibrium
phase transition a  correlation volume at $T_c$ does not evolve into a
large number of correlated zero temperature pions.
This is not encouraging.

We pause here for an aside.  The reader may be wondering why, when
the seemingly small {\it equal} quark mass $m_u = m_d = m_q$ has such
deleterious effects, we have completely neglected the difference between
the up and down quark masses.  It was noted in [\wilczek ] that unequal
quark masses allow terms of the type
$$
\Delta F \propto (\delta m)^2 (\sigma^2 - \pi_3^2 + \pi_1^2 + \pi_2^2 )~.
\eqn\iz
$$
If one is close enough to the critical point that this term matters, one
will discover an anisotropic fixed point rather than the symmetric
Heisenberg fixed point.  However, while the effect of a common quark mass,
namely the mass of the pion, is comparable to $T_c$, the effect of \iz\
is much smaller.  For example, the QCD contribution to the difference
in mass between the charged and neutral kaons is about 5 MeV [\kaonmass ].
Therefore, we need not worry about \iz\ in real experiments since
it is much less important than the effect of the ``magnetic field''
proportional to the common up and down quark mass.  Of course, \iz\
could be introduced and studied on the lattice.

It seems clear that if the standard scenario
for heavy ion collisions in which the
QCD plasma cools through $T_c$ while staying close to thermal
equilibrium is correct, then no correlation lengths will get long
enough for there to be any dramatic observable effects of the
phase transition.  The chiral ``transition'', like the
confinement/deconfinement
``transition,'' will be a smooth crossover.  If we were able to dial down the
quark masses and hence the pion mass, phenomena associated with
a second order transition would become more prominent.  Alas, in the
real world, unlike on a lattice, we have no such freedom.

\chapter{Non-equilibrium Phenomena in Relativistic Heavy Ion Collisions}

We are not done yet. The gloomy paragraph with which we ended the preceding
section began with a
conditional sentence.  In this section we will consider the
observable effects in heavy ion collisions if the plasma {\it does not}
stay close to thermal equilibrium through the transition.  There are
tantalizing hints in cosmic ray physics that point in this direction.
Among the zoo of high energy cosmic ray events known are a particularly
peculiar class of events called
Centauros [\centauro ].
These are events with total energy of order 1000 TeV in which many
(of order 100) charged hadrons
each with energies of a several TeV
and very few photons or electrons are seen in a cosmic ray induced shower.
In the sample of events in [\centauro ], there were 5 Centauros, representing
about 1\% of the events seen with energy of the appropriate order of
magnitude.
Centauros are peculiar because so many charged pions are observed without
any of the gammas that would indicate the decay of neutral pions.
This apparent violation of isospin invariance is puzzling, unless one
thinks of it in the language of a second order phase transition in which
these events can be interpreted as the creation of a volume of QCD
plasma in which the $\phi$ field has fluctuated throughout most of the
plasma in some direction in the $\pi_1 - \pi_2$ plane. This implies
correlation
throughout most of a volume
of plasma large enough that it becomes about 100 zero temperature pions.
We convinced ourselves
above that this could not happen if the plasma remains close to
thermal equilibrium.  Hence, these Centauro events provide a tantalizing
hint that it might be wise to consider the effects of going from the
symmetric phase above $T_c$ to the ordered phase rapidly without maintaining
thermal equilibrium.  This process, called quenching, has been much studied
in condensed matter physics.

We noted in the previous section that if the plasma stays close to
equilibrium, one finds below $T_c$ that the energy in a correlation
volume is small compared to the zero temperature pion mass.  However,
since the energy density at temperatures well above $T_c$ is much higher
than that below $T_c$, it is reasonable to hope that a quench, in which
the energy density does not decrease in a quasi-equilibrium fashion through
$T_c$, has a better chance of producing correlated volumes of plasma which
evolve into many zero temperature pions.
In this section we will begin to analyze the observable effects if relativistic
heavy ion collisions proceed via quenching.
In a real relativistic heavy ion collision, the phase transition will probably
occur by something in between a slow equilibrium process and a quench.
Only experiments can determine which description is more appropriate.

Let us begin by describing more carefully what a quench is, in
the context of a Heisenberg magnet
in greater than two dimensions with no applied magnetic field.
The system starts at equilibrium
at a temperature well above $T_c$,  fluctuating
among an ensemble of configurations with short correlation lengths.
One then imagines turning the temperature instantaneously to zero.
This means that the equilibrium configuration is now an ordered state
with the field aligned throughout space.  However, this is not the
configuration in which the system finds itself.  It is in one configuration
from the ensemble appropriate to a high temperature.  This configuration
then evolves according to the zero temperature equations of motion ({\it i.e.}
microcanonically, with no thermal fluctuations.)  In a condensed matter
system, the appropriate equations of motion are those of Section 6, and in
particular they are not Lorentz invariant.  What is found [\bray ]
is that
the size of correlated domains grows with time in such a way that
after a brief initial period
the correlation function has the simple scaling form
$$
C(r,t) \equiv \langle \phi (r,t) \phi (0,t) \rangle = g \bigl ( r/L(t) \bigr )
{}~.
\eqn\ja
$$
The characteristic domain size $L$ increases with time according to
$L(t) \sim t^p$, where the exponent $p$ depends only on $d$, $N$, and the
dynamic universality class.  It is important to note that the scaling
behaviour \ja\ is obtained regardless of the initial configuration.
Hence,
it is not actually necessary that the initial configuration in a
simulation (or in a heavy ion collision) be
selected from a high temperature thermal ensemble.
Any disordered initial configuration evolves to
the critical behaviour \ja\ .  This phenomenon is called self-organized
criticality.
It is also crucial for us that the domain size is not related
to an equilibrium correlation length, and in particular that in an infinite
system it grows without bound.

Let us now consider how the physics of quenching may be applied in
relativistic heavy ion collisions.
If the collision is energetic enough to create a region of plasma well
above $T=T_c$, the $\phi$ field will indeed be fluctuating among an ensemble
of disordered configurations.  At the end of the process, one certainly
has zero temperature.   The question is what happens to the $\phi$ field
in between.  One idealized possibility which has been considered by
many before us and which we considered in the previous section is that
the system stays arbitrarily close to thermal equilibrium.  Another
idealization is that thermal fluctuation ceases instanteously.
In a real collision, the plasma is cooling, and
so cannot be exactly in thermal equilibrium.  If it cools fast enough,
the configuration of the $\phi$ field will ``lag,'' and
as in a quench the system will
find itself in a configuration that is more disordered than the
equilibrium configuration appropriate for the current temperature.
However, it is clear that a real collision will not be an ideal quench,
as thermal fluctuation will not cease instantaneously. Thus,
a real collision fits in neither idealized category,
but is somewhere in between.
To the second idealization, the quench, we now turn.

There are several differences between a quench in the system of interest
to us and in the condensed matter system we described above.  First,
at zero temperature we must have a Lorentz invariant field theory, with
different equations of motion than  those used in the condensed matter
system.  While the equations of motion of section 6 describe the Lorentz
non-invariant dynamics of the order parameter near $T=T_c$, they are
inappropriate for the $T=0$ dynamics of a quench.
We propose to use the zero temperature linear sigma model
with only pion and sigma fields.  Since we will be considering energies
well below the rho and nucleon masses, we need not include these
degrees of freedom.  The sigma could also be left out, if it were not
for the fact that the appropriate initial conditions are a disordered
state in which the pion and sigma fields are equivalent up to the
effect of the bare quark masses.   The second difference is that
unlike in the magnet, the plasma in a relativistic  heavy ion collision
is expanding.  This means that the description in terms of configurations
of the field $\phi$ will not be appropriate forever.  At some time,
the energy density drops low enough that one has individual pions
flying off towards the detectors.  The third difference is that
unlike in the condensed matter systems considered in [\bray ],
we must include the effects of the bare quark masses.  As we saw
before, these correspond to a significant magnetic field.  We propose
that quenching of a 4 component Heisenberg model with the three modifications
we have mentioned be considered an idealized model for a heavy ion collision.

It is fortunate that the scenario we have just outlined, with the
exception of the significant magnetic field, is exactly the
scenario considered by Turok and Spergel [\turgel ]
as a cosmological model for large scale structure formation in the
early universe.  They study the evolution of an $O(N)$ sigma model
in an expanding universe.  They find an exact scaling solution for
the non-linear sigma model in the large $N$ limit, and do numerical
simulations for the linear sigma model for $N=4$ and $N=10$.
The main reason we can apply their results while we cannot use
those of [\bray ] is that Turok and Spergel use the Lorentz invariant
equations of motion appropriate for our problem.
They find that
the size of correlated domains, $L(t)$, grows at the
speed of light!
We are currently [\inprogress ]
extending their simulations to include
the effects of a magnetic field, and to vary the expansion rate.
The magnetic field will qualitatively change the scenario.  Instead
of having correlated domains with the $\phi$ field pointing in arbitrary
directions on the 3-sphere, at late times the $\phi$ field will be
oscillating about the sigma direction.  However, there will still be
domains in which the oscillations are in different directions.
If in a heavy ion collision the
size of these domains grows with the speed of light as it does in our
simulations [\inprogress ], the phenomenological consequences
for heavy ion collisions are dramatic.  Of course the description we
are using will only be valid for a short time. (Bjorken [\bjorken ]
estimates that a hydrodynamic description will be valid for about
10 fm/c.)  Since the plasma will be expanding slower than the speed of
light, even in this short time domains which expand at the speed of light
will grow to encompass large fractions of the total volume.  We therefore
propose that if heavy ion collisions can be modelled as a quench,
this will be detected by observing clusters of
pions in which all the pions in a region of rapidity
are correlated in internal space.
In some clusters,
there will be only charged pions; in others, only neutral ones; and in
all, charged and neutral pions will occur in some fixed ratio.

We can estimate the probability distribution of the ratio $R$ of
the number of neutral pions to the total number of pions in a correlated
region.
Let us assume that the
$\phi$ field in the region is initially equally likely to be pointing
in any direction on the 3-sphere.  This assumption may not be strictly true
because the magnetic field selects a preferred sigma direction even at
high temperatures.  However, we make the assumption in order to get a
simple analytical result.
Where $\phi$ starts will determine
in which direction it ends up oscillating about
the sigma direction.   We define angles on the 3-sphere according
to
$$
\bigl ( \sigma , \pi_3 , \pi_1 , \pi_2 \bigr ) = \bigl ( \cos \theta ,
\sin \theta \cos \phi , \sin \theta \sin \phi \cos \eta ,
\sin \theta \sin \phi \sin \eta \bigr )~.
\eqn\jb
$$
Then the ratio $R$ is given by
$$
R \equiv { n_{\pi^0} \over n_{\pi^0} + n_{\pi^+ \pi^-} } =
{ \sin ^2 \theta
\cos ^2 \phi \over
\sin ^2 \theta (\cos ^2 \phi + \sin ^2 \phi ) } = \cos ^2 \phi ~.
\eqn\jc
$$
Under the assumption that all initial values on the 3-sphere are
equally probable,
the probability distribution ${\cal P}(R)$ is determined by
$$
\int_{R_1}^{R_2}{\cal P}(R)dR = {1 \over  \pi ^2}
\int_0^{2\pi} d\eta
\int_0^{\pi} d\theta \sin^2 \theta
\int_{\arccos ( \sqrt{R_2} )}^{\arccos ( \sqrt{R_1} )} d\phi
\sin \phi
\eqn\jd
$$
and turns out to be simply
$$
{\cal P}(R) = {1 \over 2} R^{-1/2} ~.
\eqn\je
$$
Equivalently, the probability that $R<R_1$ is given by $\sqrt{R_1}$.
If heavy ion collisions are described by a quench, there should
be large regions of the collision volume containing clusters of pions
in which $R$ is constant, and
the values $R$ takes in different such regions should be distributed according
to \je\ .

As one application of \je\ , we note that the probability that the neutral
pion fraction $R$ is less than $.01$ is $0.1$!  This is a graphic illustration
of how different \je\ is from what one would expect if individual pions
were independently randomly distributed in isospin space.  It also makes
Centauro events in which less than 1\% of the outgoing particles from a
heavy ion collision are neutral pions
seem much less surprising than they first appeared.
The analysis of the Centauro data is difficult for several reasons.
Most important of these is the limitation imposed by small statistics.
Also, if a Centauro event occurs too high above
the detector, so many secondary photons will be produced that the
event will not be recognized as a Centauro.
Third, in a Centauro event all of the particles from the collision strike
a small region  of the detector and it is impossible to isolate the
central rapidity region.  This combined with the fact that the detectors
do not distinguish between charged pions and charged baryons
has several unfortunate consequences.  It means that events in which
there are two or more correlated clusters of pions are not detected
as Centauros.  Only those with a single cluster are so identified.
Also, the opposite type of event in which $R$ is close to 1, will
not have a dramatic signature since there will always be charged
baryons present from the two initial nuclei.
For all these reasons, we feel it is impossible to extract
a meaningful probability distribution ${\cal P}(R)$ from the cosmic
ray data.

When relativistic heavy ion collisions occur at high enough energies
in a laboratory colliding beam facility, all of the difficulties
of the cosmic ray experiments will be rapidly overcome.  That will
be the time to look for correlated clusters of pions, and to
look for a distribution like \je\ , and hence to determine whether
these collisions proceed by a process close to the idealized quench
we have considered here.

\chapter{Conclusions}

The future of the study of the QCD phase transition
on the lattice looks promising.
As simulations improve, they will begin
to investigate the plethora of static critical phenomena we discussed
in the first sections of this paper.  Critical exponents, the equation
of state, the critical behaviour of the pion and sigma susceptibilities,
$\rho$ and $A_1$ correlation functions, and tricritical exponents are all
out there waiting to be measured.

Because heavy ion collisions are dynamical processes, there are more
possible scenarios and the situation is not clear cut.  We have
considered two idealized models.  If the cooling plasma stays
arbitrarily close to thermal equilibrium, our conclusions are
disappointing.  Because the pion is so heavy compared to $T_c$, correlation
lengths will not become particularly long.  On the other hand,
in the other idealized model we considered, in which thermal fluctuations
are rapidly quenched and the cooling plasma is in a sense maximally
out of thermal equilibrium, dramatic phenomena are possible.
Correlated volumes will form, and their size will  not be determined
by any equilibrium correlation length.  Indeed they may grow at the speed
of light until hadronization occurs.  This will have the consequence
that clusters of large numbers of pions will be detected in which
the ratio of neutral to charged pions will be constant.  This ratio
will be different in different clusters, and will follow a probability
distribution which is skewed towards having few  neutral pions.
We eagerly await the verdict of experiment as to which scenario is more
appropriate.

\ack{We are grateful for the fruitful discussions we had with Bert Halperin.
We both also acknowledge the hospitality of Harvard University, where
part of this work was completed.}

\appendix{}

In this appendix, we reproduce the equation of state to order $\epsilon^2$
[\bww ], and give various other results used in section 3.

The equation of state is found by doing an $\epsilon$ expansion of the relation
$$
\langle \tilde \sigma \rangle = 0 ~,
\eqn\za
$$
where $\tilde \sigma = \sigma - M$.  The resulting expansion can be
expressed in terms of the variables $y \equiv H/M^\delta$ and
$x\equiv t/M^{1/\beta}$, where
$$
{1 \over \beta} = 2 + {6 \over N + 8}\epsilon + 4 {(N+5)(7-N) \over
(N+8)^3 } \epsilon^2 + O(\epsilon^3)
\eqn\zb
$$
and
$$
\delta = 3 + \epsilon + {N^2 + 14N + 60 \over 2(N+8)^2}\epsilon^2
 + O(\epsilon^3)~.
\eqn\zc
$$
We are interested in $\epsilon = 1$ and $N=4$.
We choose to measure fields in units such that $y=1$ at $t=0$,
and $x=-1$ at the coexistence curve ($H=0, ~ t<0$).
The equation of state is
$$
y=f(x)~.
\eqn\zd
$$
The function $f(x)$ is given by
$$
\eqalign{
f(x) &=~ 1~ +~ x \cr
     &+~ {\epsilon \over 2(N+8)}\Bigl( 1 + {\epsilon \over 2(N+8)}
\bigl[ N-1+6\ln 2 -9\ln 3 + (N-1)\ln (x+1) \bigr] \Bigr) \cr
     &~~~~\times \Bigl( 3(x+3)
\ln (x+3) + (N-1)(x+1)\ln (x+1) + 6x\ln 2 - 9(x+1)\ln 3 \Bigr)  \cr
     &+~\Bigl( {\epsilon \over 2(N+8) } \Bigr) ^2 \Bigl\lbrace ~ {1 \over 2}
(10-N)(x+1)\bigl[\ln ^2(x+3) - \ln^2 3 \bigr] \cr
     &~~~~+ 36\bigl[ \ln^2 (x+3)
-(x+1)\ln^2 3 + x\ln^2 2 \bigr] \cr
     &~~~~-54\ln 2 \bigl[ \ln (x+3) + x \ln 2 - (x+1)\ln 3 \bigr]
+3(N-1)\bigr( \ln {27\over 4}\bigl) (x+1)\ln (x+1) \cr
     &~~~~+{212 + 17N - 4N^2 \over N+8}\bigl[ (x+3)\ln (x+3) + 2x\ln 2
-3(x+1)\ln 3 \bigr] \cr
     &~~~~+ (N-1)(x+1)\ln (x+1) \ln (x+3)-{N\over 2}(N-1)(x+1)\ln^2 (x+1) \cr
     &~~~~+ {N-1 \over N+8}(19N+92)(x+1)
\ln (x+1) - 2(N-1)\bigl[ (x+6)I_1(\rho) - 6(x+1)I_1(3/4) \bigr] \cr
     &~~~~-6(N-1)\bigl[I_2(\rho)-(x+1)I_2(3/4)\bigr] + 4(N-1)\bigl[
I_3(\rho) - (x+1)I_3(3/4) \bigr] \Bigr\rbrace + O(\epsilon^3) \cr}
\eqn\ze
$$
where
$$
\rho \equiv {x+3 \over 4(x+1)}
\eqn\zf
$$
and
$$
\eqalign{
I_1(\rho) &\equiv \int_0^\rho {du \ln u \over u(1-u)}\bigl[ \sqrt{1-u/\rho}
-1 \bigr] - \int _\rho^\infty {du \ln u \over u(1-u)} \cr
I_2(\rho) &\equiv \rho {dI_1 \over d\rho} \cr
I_3(\rho) &\equiv I_1(\rho) + 2 I_2(\rho)~~.\cr}
\eqn\zg
$$
The behaviour of $I_1(\rho)$ near the coexistence curve is given by
$$
I_1(\rho) \sim {1\over 4\rho}\bigl( \ln^2 4\rho + 2\ln \rho \bigr)~,~~\rho
\rightarrow \infty ~.
\eqn\zh
$$
The leading terms of $f(x)$ for large $x$ are
$$
f(x) \sim \Bigl( 1+ {3\ln (4/27) \over2(N+8) }\epsilon + O(\epsilon^2)~
\Bigr)~x^\gamma~,~~x\rightarrow \infty
\eqn\zi
$$
where
$$
\gamma = 1 + {N+2 \over 2(N+8)}\epsilon +{(N+2)(N^2+22N+52)\over
4(N+8)^3 }\epsilon^2 + O(\epsilon^3)~.
\eqn\zj
$$
Of course, this result for $\gamma$ and the results for $\beta$ and
$\delta$ in (A2) and (A3) are consistent with the scaling relations \db\ .
We saw in section 3 that the behaviour \zi\ determines the sigma and pion
masses
for $H=0$ and $t>0$.

In section 3, in order to determine
$m_\sigma^2$ near the coexistence curve we needed the behaviour of $f'(x)$ for
$x\rightarrow -1$.
In this region, $f(x)$ is given by [\wz ]
$$
\eqalign{
f(x) \sim (x+1)\Biggl\lbrace & 1+\epsilon \Biggl[ {N-1 \over 2(N+8)} \ln (x+1)
+{3(1+3\ln (2/3) )\over 2(N+8)} \Biggr] \cr
& +\epsilon^2 \Biggl[ {(N-10)(N-1) \over 8(N+8)^2}\ln^2 (x+1) \cr
&~~~~~+ {N-1 \over 4(N+8)^2}
\Bigl[ N+27 +18\ln 2 - 9\ln 3 - {60\over N+8} \Bigr] \ln (x+1) \Biggr]
+O(\epsilon^3) \Biggr\rbrace~. }
\eqn\zk
$$
One next inverts \zk\ to obtain $x+1$ in terms of $y=f(x)$,
$$
(x+1) = c_1 y + \tilde c_2 y^{1-\epsilon /2} + O\Bigl(y^{2-O(\epsilon)}
\Bigr)~,
\eqn\zka
$$
differentiates
the result with respect to $x$, and obtains
$$
{\beta m_\sigma^2 \over M^{\delta -1}} \rightarrow f'(x)
= { 1 \over c_1 + c_2 y^{-\epsilon /2}} ~~{\rm for}~x\rightarrow -1~,
\eqn\zl
$$
with
$$
c_1 = {9 \over N+8}\Biggl( 1- {\epsilon \over 2(N+8)}\Bigl[ (N+8)\ln 2 -
9\ln 3 + {25N^2 +142N + 76 \over 9(N+8)} \Bigr] \Biggr) + O(\epsilon^2 )
\eqn\zm
$$
and
$$
\eqalign{c_2 &= \tilde c_2 \Bigl(1-{\epsilon \over 2}\Bigr) \cr
&={N-1 \over N+8}\Biggl\lbrace 1- {\epsilon \over 2}\Bigl( 1- {1 \over
(N+8)^2} \bigl[ 9(N+8)\ln 3 + 22N + 116 \bigr] \Bigr) \Biggr\rbrace
+ O(\epsilon^2)~~.}
\eqn\zn
$$
As $H$ is lowered to zero at fixed $t<0$, $m_\sigma^2$ at first tends
toward a constant,
and then goes to zero according to $m_\sigma^2 \propto H^{\epsilon /2}$
when the $c_2$ term takes over from the $c_1$ term.  For $N=4$, this occurs
when
$$
{H \over M^\delta } \lsim \Bigl( { c_2 \over c_1} \Bigr)^{2/\epsilon}
\sim \Bigl( {1\over 3} \bigl[1+0.96\epsilon\bigr] \Bigr)^{2/\epsilon}
\sim 0.4 {\rm ~for~} \epsilon = 1~.
\eqn\zo
$$
Of course, the numerical value for $\epsilon =1$ should not be taken
too seriously.  The qualitative result is clear nevertheless.

We end this appendix by describing how the figures in section 3 were
obtained.  When evaluated at $\epsilon =1$, the expression \ze\ for
$f(x)$ has several problems.  First, at large $x$ it does not grow as
$x^\gamma$.  Rather, it increases like $x\ln^2 x$.
Also, for $x\rightarrow -1$,
$f(x)$ given by \ze\ does not satisfy \zka\ .  In fact, for $x \lsim -0.95$,
$f(x)<0$ which is unphysical.  Both of these problems arise because we are
setting $\epsilon =1$ in
an expansion of $f(x)$ to order $\epsilon^2$ which is valid for $\epsilon
\rightarrow 0$.  In order to illustrate the correct qualitative
behaviour, we constructed a function $f(x)$ which smoothly interpolates
between \zi\ at large $x$, \zka\ near $x=-1$, and \ze\ in between.
Using this function $f(x)$, we obtained Figure 1 by solving
$H/M^\delta=f(t/M^{1/\beta})$ for $M$ at various values of $t$ and $H$.
We then calculated the results for $m_\pi^2$ and $m_\sigma^2$ shown
in Figures 2 and 3 using \ef\ and \efc\ .
Because of the limitations imposed by working at $\epsilon =1$, all three
figures should be viewed as illustrations of qualitative behaviour.

\vfill
\eject

\FIG\zza{The order parameter $M$ as a function of reduced temperature
$t\equiv (T-T_c)/T_c$
for magnetic fields $H=$ 0, 0.002, 0.005, and 0.02.
Like $t$, both $M$ and $H$ are dimensionless. They are obtained from their
dimensionful counterparts by dividing by non-universal dimensionful constants
defined in such a way that
for $t<0$ and $H=0$ the order parameter is given by $M=(-t)^\beta$, and
for $t=0$ it satisfies $M=H^{1/\delta}$.  }
\FIG\zzb{$m_\pi^2$ (Figure 2a) and $m_\sigma^2$ (Figure 2b)
as functions of $t$ for
$H=$ 0, 0.002, 0.005, and 0.02.  Since $M$ and $H$ are in dimensionless
scaled units, so are $m_\pi^2$ and $m_\sigma^2$.
For $t=0$, $m_\pi^2 = H^{(\delta -1)/\delta}$ and $m_\sigma^2 = \delta
m_\pi^2$.
For $H=0$ and $t>0$ (and for large enough $t$ for any $H$) $m_\sigma^2 =
m_\pi^2 \sim t^\gamma$.  For $t<0$ and $H\rightarrow 0$, $m_\pi^2 \sim H$
and the sigma mass decreases to zero as shown in Figure 3.}
\FIG\zzc{$m_\sigma^2$ as a function of $H$ for $t=-0.2$.
For large $H$ it behaves as if it will be non-zero at $H=0$, but in fact
for $H\rightarrow 0$ it decreases like $m_\sigma^2 \sim H^p$ where to lowest
order in $\epsilon$, $p=\epsilon /2 = 1/2$.}

\endpage
\refout
\endpage
\figout
\end